\documentclass[letterpaper, 10 pt, conference]{ieeeconf}  

\IEEEoverridecommandlockouts                              

\usepackage{graphics} 

\usepackage{xcolor}
\let\labelindent\relax
\usepackage{framed,enumitem} 
\usepackage{epsfig} 
\usepackage{times} 
\usepackage{amsmath, bm}
\usepackage{amssymb}  
\usepackage{arydshln}
\usepackage{amsfonts}
\usepackage{algorithm}
\usepackage[noend]{algpseudocode}
\usepackage[noadjust]{cite}
\usepackage{stfloats}
\usepackage{bm}

\newcommand{\YB}[1]{{\color{black}#1}}

\title{\LARGE \bf
\YB{On Improved Commutation for Moving-Magnet Planar Actuators*}
}

\author{Yorick Broens, Hans Butler and Roland T\' oth
\thanks{*This work has received funding from the ECSEL Joint Undertaking (JU) \YB{programme} under grant agreement No 875999 and from the European Union within the framework of the National Laboratory for Autonomous Systems (RRF-2.3.1-21.2022-00002).}
\thanks{\YB{Y. Broens}, \YB{H. Butler} and \YB{R. T\'oth} are with the Department of Electrical Engineering, Eindhoven University of Technology, Eindhoven, The Netherlands. \YB{H. Butler} is also affiliated with ASML, Veldhoven, The Netherlands. \YB{R. T\'oth} is also affiliated with the Systems and Control Laboratory, Institute for Computer Science and Control, Hungary,  ({\tt  email: Y.L.C.Broens@tue.nl}). }
}

\begin{document}

\maketitle
\thispagestyle{empty}
\pagestyle{empty}

\begin{abstract}
The demand for high-precision and high-throughput motion control systems has increased significantly in recent years.
The use of moving-magnet planar actuators (MMPAs) is gaining popularity due to their advantageous characteristics, such as complete environmental decoupling and reduction of stage mass. Nonetheless, model-based commutation techniques for MMPAs are compromised by misalignment between the mover and coil array and mismatch between the ideal electromagnetic model and the physical system, often leading to decreased system performance. To address this issue, a novel improved commutation approach is proposed in this paper\YB{, which is applicable for general planar motor applications,} by means of dynamic regulation of the position dependence of the ideal model-based commutation algorithm,
which allows for attenuation of magnetic misalignment, manufacturing inaccuracies and other unmodelled phenomena. The effectiveness of the proposed approach is validated through experiments using a state-of-the-art moving-magnet planar actuator prototype.

\end{abstract}

\section{Introduction}
\label{Section:Introduction}
In recent years, the demand for high-precision and high-throughput motion control systems has seen a significant increase across various fields, such as microelectronics, biotechnology, and nanotechnology, see \cite{zhang2016development,qu2016motion,Butler}. To achieve highly accurate positioning of the mover while allowing for increased throughput, new electromagnetic actuator configurations with improved performance have been developed, particularly in the form of planar motors utilizing a moving-magnet configuration, see \cite{Proimadis-phd,Rovers-phd,Lierop-phd}. The use of \emph{moving-magnet planar actuators} (MMPAs) is gaining popularity due to their advantageous characteristics, such as complete environmental decoupling of the mover and reduction of the stage mass, over their alternatives. However, these benefits come at the cost of introducing an additional surge of complexity from a motion control perspective\YB{,} due to the presence of complex nonlinear multi-physical effects that can only be approximately modeled based on first-principles knowledge, see \cite{6557499}. Additional complexity arises from position dependent effects which are introduced by relative position measurements and actuation of the moving-body. To address these, coordinate frame transformations are required which accurately connect the actuation forces (stator frame) and position measurements (metrology frame) to the specific point of control on the moving-body (translation frame), see \cite{Steinbuch2013}.

Typically, motion control design for MMPAs is simplified through the use of model-based commutation approaches, see \cite{Proimadis-phd,Rovers-phd,6557499,Lierop-phd}. These approaches rely on a first-principles based model of the inverse \emph{electromagnetic} (EM) behavior of the motor, with the goal of \YB{eliminating} the nonlinear electromagnetic interactions between the coil array and the magnet array, thereby allowing for independent \YB{linear} control of the mechanical \emph{degrees of freedom} (DoFs) of the mover. However, despite the benefits of these approaches, they still require a precise characterization of the highly nonlinear \YB{EM} behavior of the actuator, which can be difficult to obtain with sufficient accuracy due to the complex nonlinear multi-physical \YB{nature of the involved phenomena} combined  with unknown manufacturing inaccuracies.
Additionally, model-based commutation techniques utilized in magnetically levitated movers\YB{, see \cite{Proimadis-phd,Rovers-phd,Lierop-phd}}, encounter a challenge of misalignment during system initialization due to position uncertainty of the magnetic-mover with respect to the coil array, leading to a decrease in system performance. These properties necessitate the adoption of sophisticated calibration strategies for accurate alignment of the moving-magnets with the stator to allow for high-precision positioning of the moving-body. 
\YB{Various efforts have been made to mitigate these effects using learning-based feedforward techniques, see \cite{proimadis2021learning,9659174}. However, the substantial data requirements for accurately capturing the complex EM relationships inherent to MMPAs pose a significant hurdle.  To address these issues, this paper introduces novel approaches to enhance the existing state-of-the-art commutation.} First, a static gradient-descent based optimization approach is \YB{introduced} which provides automatic calibration of the commutation frame in terms of aligning the magnetic-mover with the coil array without the use of additional sensors.
Secondly, an active regulatory approach is presented that can adapt the commutation frame to local variations of the EM relations due to misalignment, coil pitch, eddy currents or other manufacturing imperfections. \YB{The latter method incorporates a secondary control loop in combination with a learning-based feedforward. }

\vspace*{-1mm}
The main contributions of this paper are:
\vspace*{-2mm}
\begin{itemize}
\item[(C1)] The development of a novel electromagnetic calibration approach for MMPAs by means of a gradient-descent based optimization strategy. The proposed approach allows for compensation of static misalignment of the moving-magnets with respect to the coil array.
  \item[(C2)] The development of a novel improved commutation approach which  dynamically regulates the commutation frame, therefore attenuating for effects of misalignment, manufacturing inaccuracies and other remnant effects.
\item [(C3)] The development of a position dependent learning-based feedforward 
with the aim of improving the dynamic regulation of the commutation frame.
\end{itemize}

This paper is organized as follows. First, the problem formulation is presented in Section \ref{Section_Problem_formulation}. Next, Section \ref{Section_GRADDESC} presents the proposed gradient descent based static calibration of the commutation frame. Section \ref{Section_Dynamic_Regulation} introduces
the proposed dynamic regulation of the commutation frame. In Section \ref{Section:CommutationFeedforward}, the design of a learning-based commutation feedforward is proposed for \YB{enhancing} the dynamic regulation. Section \ref{Section_Experimental_Validation} presents experimental results of the proposed approaches on a state-of-the-art MMPA prototype. Finally, Section \ref{Section_Conclusions} presents the \YB{overall} \YB{conclusions} on the presented work.


\vspace{-1mm}
\section{Problem formulation}
\label{Section_Problem_formulation}
\subsection{Background}

\begin{figure}[t]
\vspace{7pt}
    \centering
    \includegraphics[trim={.5cm 0.5cm 2.2cm 0cm}
    ,width=.95\linewidth]{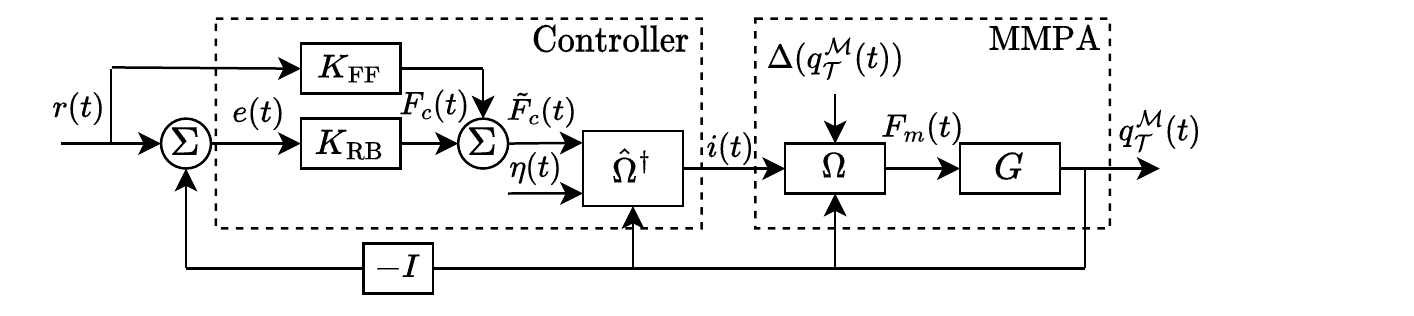}
    \caption{\YB{Schematic representation of the control structure of an MMPA system, where $\Delta \big(q_{\mathcal{T}}^{\mathcal{M}}(t)\big )$ describes the variations of the EM relationship due to misalignment and remnant EM effects, while ${\eta}(t)$ represents an additional control parameter which allows for active commutation control.}}
    \label{Fig:ContolInterconnection}
    \vspace{-4mm}
    \vspace{-7pt}
\end{figure}

The dynamic behavior of an MMPA system is governed by a combination of electromagnetic and mechanical phenomena, see Figure \ref{Fig:ContolInterconnection}, resulting in a complex \emph{multiple-input multiple-output} (MIMO) system, which exhibits position dependent effects due to the relative displacement of the mover with respect to the measurement (metrology) and actuation (stator) frames \YB{\emph{assuming a rigid-body (RB) mover}}.
The equations of motion are given by:
\vspace{-1mm}
\begin{equation}
M\ddot{q}_\mathcal{T}^{\mathcal{M}}(t) + D\dot{q}_\mathcal{T}^{\mathcal{M}}(t)+Kq_\mathcal{T}^{\mathcal{M}}(t) =H F_m(t)
    \label{eq:MechanicalModel_mover}
\end{equation}

\vspace{-1mm}
\noindent where $M$,$D$ and $K$ are the symmetric mass, damping and stiffness matrices of the mover and $q_\mathcal{T}^{\mathcal{M}}(t) \in \mathbb{R}^{n_q}$ corresponds to the position vector of the mover in the metrology coordinate frame.
$H\in \mathbb{R}^{n_q \times n_{F}}$ represents the mapping of the \YB{physical} forces \YB{$F_m(t)$} acting on the magnet plate \YB{at} its center of gravity. \YB{The} EM interaction, which relates the currents in the stator coils\YB{, $ i(t) \in \mathbb{R}^i$,} to forces $F_m(t)$ that are exerted on the magnet plate,
is given by (see \cite{Rovers-phd}):
\vspace{-1mm}
\begin{equation*}
F_m(t) =\Omega \left(q_{\mathcal{T}}^{\mathcal{S}}(t) \right)i(t),
     \label{eq:EMInteraction}
\end{equation*}

\vspace{-1mm}
\noindent where $q_\mathcal{T}^{\mathcal{S}}(t) \in \mathbb{R}^{n_q}$ corresponds to the position vector of the mover in the stator coordinate frame.
In case of misalignment between the two coordinate frames, i.e. $q_{\mathcal{T}}^{\mathcal{S}} = q_{\mathcal{T}}^{\mathcal{M}} + \Delta(q_{\mathcal{T}}^{\mathcal{M}})$, 
the EM interaction \YB{is} reformulated as:
\vspace{-1mm}
\begin{equation}
F_m(t) = \Omega \left(q_{\mathcal{T}}^{\mathcal{M}}(t) + \Delta \big(q_{\mathcal{T}}^{\mathcal{M}}(t)\big) \right)i(t),
     \label{eq:EMInteraction}
\end{equation}


\vspace{-1mm}
\noindent where $\Delta(q_{\mathcal{T}}^{\mathcal{M}}(t))$ represents the variations of the EM relationship due to misalignment, coil pitch, eddy currents and other manufacturing imperfections \YB{and it is unknown}.
Furthermore, by combining (\ref{eq:MechanicalModel_mover}) and (\ref{eq:EMInteraction}), the MMPA system can be represented in state-space form, denoted by $\mathcal{G}$, as:
\vspace{0mm}
\YB{\begin{equation}
    \mathcal{G}:= \begin{cases}
            \begin{pmatrix}
                \dot{x}(t) \\ q_{\mathcal{T}}^{\mathcal{M}}(t)
            \end{pmatrix}
            =
            \begin{pmatrix}
                \mathcal{A} & \mathcal{B}(q_{\mathcal{T}}^{\mathcal{M}}(t)) \\ 
                \mathcal{C} & 0
            \end{pmatrix} \begin{pmatrix}
                {x}(t) \\ i(t)
            \end{pmatrix},
    \end{cases}
    \label{eq:PlantDynamics}
\end{equation}}
    \\

\vspace{-5mm}

\noindent \YB{where $x(t)\in \mathbb{R}^{2n_q}$ corresponds to the state vector. Note that for constant ${\tt q} \equiv q_\mathcal{T}^{\mathcal{M}}(t)$, the dynamics expressed by $\mathcal{A},$ $\mathcal{B}({\tt q})$, $\mathcal{C}$ from (\ref{eq:PlantDynamics}) are linearized around a forced equilibrium position.
A collection of these is often referred to as the \emph{local dynamics} of (\ref{eq:PlantDynamics}) for a \emph{frozen position} across the operating envelope of the system, each denoted by $\mathcal{G}(\tt q)$.}



It is a common practice to aim for independent control of the mechanical DoFs of an MMPA. To achieve this, rigid body decoupling methods as described in \cite{STEINBUCH1998278} are generally

\noindent 
utilized. 
Actuator decoupling in MMPA systems is achieved through a model-based commutation algorithm\YB{, denoted by $\hat{\Omega}^\dagger \big(q_{\mathcal{T}}^{\mathcal{M}}(t) \big)$}, which is \YB{extensively} discussed in \cite{Proimadis-phd,Rovers-phd,Lierop-phd}. This algorithm aims to eliminate the input non-linearity of the system based on the inverse of the ideal EM relationship, i.e.\begin{small} $\Omega \big(q_{\mathcal{T}}^{\mathcal{M}}(t) \big)$$\hat{\Omega}^\dagger  \big(q_{\mathcal{T}}^{\mathcal{M}}(t) \big) = I$ \end{small}. However, due to the presence of 
\YB{$\Delta(q_{\mathcal{T}}^{\mathcal{M}})$}, the highly nonlinear behavior of the motor may not be perfectly mitigated, resulting in a mismatch between the control forces\YB{, i.e. $\Tilde{F}_c(t)=\hat{\Omega}\big(q_{\mathcal{T}}^{\mathcal{M}}(t) \big)i$}, and the physical forces acting on the mover \YB{$F_m(t)$, given by (\ref{eq:EMInteraction})}. 

We propose an approach that involves an extension of the \YB{current state-of-the-art control configuration for MMPAs, which is comprised of a model-based commutation algorithm $\hat{\Omega}^\dagger (q_\mathcal{T}^{\mathcal{M}}(t))$, a rigid-body feedback controller $K_{\mathrm{FB}}$ and a rigid-body feedforward controller $K_{\mathrm{FF}}$ (\cite{Proimadis-phd}). By incorporating the controllable parameter ${\eta}(t)$, our objective is to:}
\begin{equation}
\resizebox{.91\hsize}{!}{$
\min_{\eta(t)}
\left 
\| \Omega\Big(q_{\mathcal{T}}^{\mathcal{M}}(t)+\Delta(q_{\mathcal{T}}^{\mathcal{M}}(t))\Big) \YB{\hat{\Omega}}^\dagger \Big (q_{\mathcal{T}}^{\mathcal{M}}(t)+\eta(t)\Big) - I \right \|,
\label{costfunctionpar2}$}
\end{equation}

 \noindent \YB{at every time instant, }such that effects originating from \YB{$\Delta(q_{\mathcal{T}}^{\mathcal{M}}(t))$}
 are attenuated for, allowing high-precision motion control of the mover.


\vspace*{-1mm}
\subsection{Problem statement}

\label{Subsection_problem_statement}

The problem that is being addressed in this paper is to develop an enhanced commutation approach which aims at eliminating the effects of $\Delta(q_{\mathcal{T}}^{\mathcal{M}} (t))$ through regulation of $\eta(t)$, such that the cost function expressed by (\ref{costfunctionpar2}) is minimized. We aim at developing an improved commutation approach, such that the following requirements are satisfied:
\begin{itemize}
  \item[(R1)] The approach is able to establish an EM calibration of the commutation frame using local measurements, thus avoiding 
  extensive remodelling of the EM relations.
  \item[(R2)] The approach is capable to attenuate for EM discrepancies, originating from coil pitch, eddy currents or other manufacturing imperfections.
   \item[(R3)] The closed-loop system is stable during \YB{calibration and dynamic regulation of $\eta(t)$}.
\end{itemize}

\vspace*{-1mm}
\section{Static calibration of the commutation}
\label{Section_GRADDESC}

In this section, a static EM calibration method for the commutation frame is introduced, which involves aligning the magnetic-mover with the coil array without relying on supplementary sensors. \YB{For this purpose, optimization of a static $\eta$ is accomplished} to address commutation frame misalignment under the assumption that $\Delta$ is invariant with respect to $q_{\mathcal{T}}^{\mathcal{M}}$. 
\begin{algorithm}[b]
\caption{Gradient-descent based static calibration}\label{euclid}
\begin{algorithmic}[1]
\State Set initial step size $\lambda_{k}$
\For {$k=1:n_{\mathrm{iterations}}$}
\For {$i=1:n_{\tt {p}}$} 
\State  Measure $\|F_c(\tt q_i+{\eta}_k)\|_2$ 
\For {$j=1:n_{{\eta}}$}
\State Measure $\|F_c(\tt q_i+{\eta}_k +\xi_j )\|_2$ 
\EndFor
\State \textbf{end}
\State Construct $\nabla \|F_c(\tt q_i+{\eta}_k)\|_2$  by (\ref{Gradientestimation})
\EndFor
\State \textbf{end}
\State Construct $\eta_{k+1}$ by (\ref{UpdateLaw})
\State update $\lambda_{k+1}$
\EndFor
\State \textbf{end}
\end{algorithmic}
\end{algorithm} 
As the model-based commutation relies on the inverse of the ideal EM behavior of the motor, a static misalignment of the commutation frame results in static disturbance forces \YB{as seen by the rigid body feedback controller $K_{\mathrm{FB}}$}, which appear in a position dependent manner due to the position dependent nature of \YB{$\Omega(q_{\mathcal{T}}^{\mathcal{M}}+\Delta)$}. These disturbance forces can be indirectly measured from the steady-state response of the rigid body feedback control forces \YB{$F_c(t)$, see Figure \ref{Fig:ContolInterconnection}}, as the integral action of the controller \YB{builds up a constant force component to attenuate these effects.} In this context, the static attenuation forces can be viewed as a function of the position, i.e. $F_c(q_{\mathcal{T}}^{\mathcal{M}}(t)+\eta)$, which can be \YB{minimized over the operating envelope of the system} through \YB{$\eta$ by approximating $\Delta$. By considering local measurements, e.g. ${\tt q} \equiv q_\mathcal{T}^{\mathcal{M}}(t)$, $\Delta$ can be approximated by:}
\vspace*{-1.5mm}
\begin{equation}
\min_{{\eta}}\sum_{i=1}^{n_{\mathrm{p}}} \|  F_c \left ({\tt q_i}+\eta\right )   \|_2,\label{GD_ObjectiveFunction}
\end{equation}

\vspace*{-1.5mm}

\noindent where $n_{\mathrm{p}}$ corresponds to the number of local positions. The objective function given by (\ref{GD_ObjectiveFunction}) \YB{is solved} by a \emph{gradient-descent} (GD) based setting, see \cite{ruder2016overview,berglund2022novel}:
\vspace*{-1.5mm}
\begin{equation}
\resizebox{.91\hsize}{!}{$
    \Theta{\eta}_{k+1} = \Theta{\eta}_k - \lambda_k \begin{pmatrix}
      \nabla \left| \left |F_c({\tt q_1}+{\eta}_k) \right | \right |_2^\top  \ \hdots \ \nabla \left| \left |F_c({\tt q_{n_{\tt p}}}+{\eta}_k) \right | \right |_2^\top 
    \end{pmatrix}^\top ,$}
\end{equation}

\vspace*{-1mm}
\noindent where $k \in \mathbb{N}$ corresponds to the optimization step, $\lambda_k$ is the learning rate and \begin{small}$ \Theta  = ( I_{{n_{\mathrm{p}}}\times 1} \otimes I_{n_{{\eta}} \times n_{{\eta}}})$\end{small} denotes a projection matrix. The corresponding GD optimizer is given by:
\vspace*{-1.5mm}
\begin{equation}
\resizebox{.91\hsize}{!}{$
    {\eta}_{k+1} = {\eta}_k - \lambda_k\Theta^\dagger \begin{pmatrix}
      \nabla \left| \left |F_c({\tt q_1}+{\eta}_k) \right | \right |_2^\top  \ \hdots \ \nabla \left| \left |F_c({\tt q_{n_{\tt p}}}+{\eta}_k) \right | \right |_2^\top 
    \end{pmatrix}^\top$} 
    \label{UpdateLaw}
\end{equation}

\vspace*{-1mm}
In practical settings, the direct measurement of gradients is infeasible, hence an online gradient approximation method is required which capitalizes on the accessible static feedback control forces. This is achieved through $(n_\eta+1)$ measurements per local position \YB{in each iteration step $k$ }by introducing a perturbation parameter $\xi_j$ with $j \in [1 \  n_{\eta}]$, thus yielding the following gradient approximation:
\vspace*{-1mm}
\begin{equation}
\resizebox{.91\hsize}{!}{$
    \nabla \|F_c ({\tt q_i}+\eta_k) \|_2 \approx 
    \left( \begin{array}{c}
  \xi_1^{-1}\left(\left | \left | F_c({\tt q_i}+{\eta}_k+\xi_1 )\right| \right|_2-  \|  F_c({\tt q_i}+{\eta}_k ) \|_2 \right)
  \\
  \vdots \\
  \xi_{n_{\eta}}^{-1}\left(\left | \left | F_c({\tt q_i}+{\eta}_k+\xi_{n_\eta} )\right| \right|_2-  \|  F_c({\tt q_i}+{\eta}_k ) \|_2 \right)
   \end{array}\right)
    $}
    \label{Gradientestimation}
\end{equation}

\vspace*{-1mm}
The resulting approach is described by Algorithm \ref{euclid}. The key concept is to obtain measurements of two types of control forces: (i) the \YB{\emph{averaged steady-state control forces}} and (ii) the \YB{\emph{averaged perturbed static control forces}} for all $n_{\tt p}$ local positions. These measurements are then used to reconstruct the local gradients, providing the calibration parameter update process ${\eta}_{k+1}$ by (\ref{UpdateLaw}). To speed up the optimization process, the step size $\lambda_{k+1}$ can be updated in an iterative manner by using line search approaches, see \cite{berglund2022novel}, or by 
considering a Gauss-Newton type approaches, e.g. using a similar strategy for Hessian approximation as (\ref{Gradientestimation}). \YB{
The proposed method fulfills (R1) through exclusive reliance on local measurements. Additionally, the calibration process produces a finite static gain adjustment, leading to algorithm convergence, see \cite{blaschke1997convergence}. Consequently, the post-calibration closed-loop control ensures stability, thereby satisfying (R3) and contributing as Contribution C1 in the paper.}





\vspace*{-1mm}
\section{Dynamic regulation of the commutation}
\label{Section_Dynamic_Regulation}
\begin{figure}[t]
\vspace{7pt}
    \centering
        \includegraphics[trim={.5cm 0.5cm 2.2cm 0cm}
    ,width=.95\linewidth]{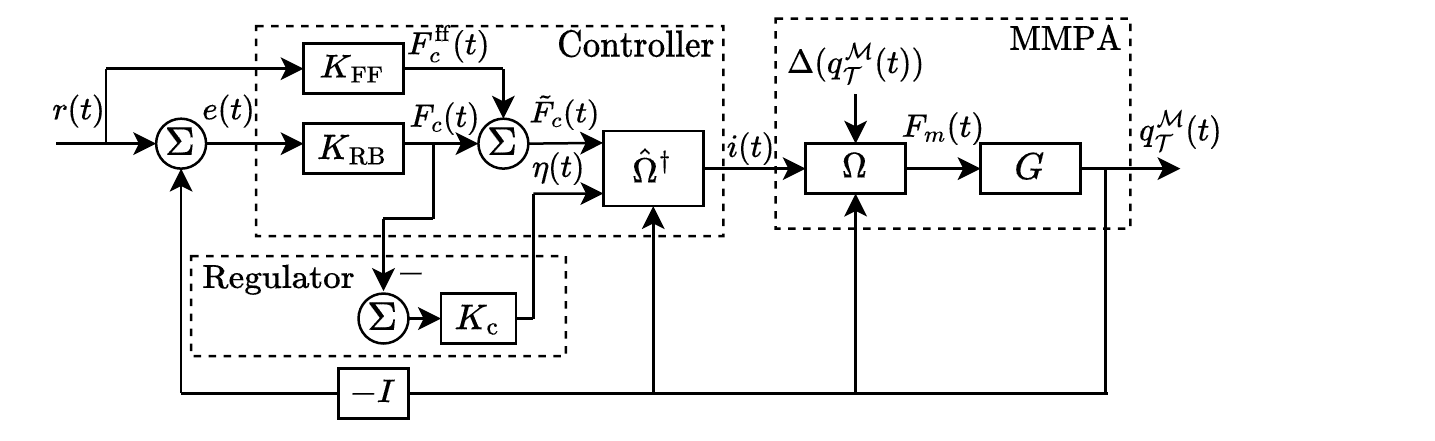}
    \caption{Proposed feedback control interconnection for dynamic regulation of the commutation frame.}  \label{Fig:ContolInterconnection1}
    \vspace*{-4mm}
    \vspace{-7pt}
\end{figure}

\begin{figure}[b]
 \vspace*{-1mm}
    \centering
\includegraphics[trim={1.9cm 0cm 2.2cm .7cm}    ,height=4cm,width=\linewidth]{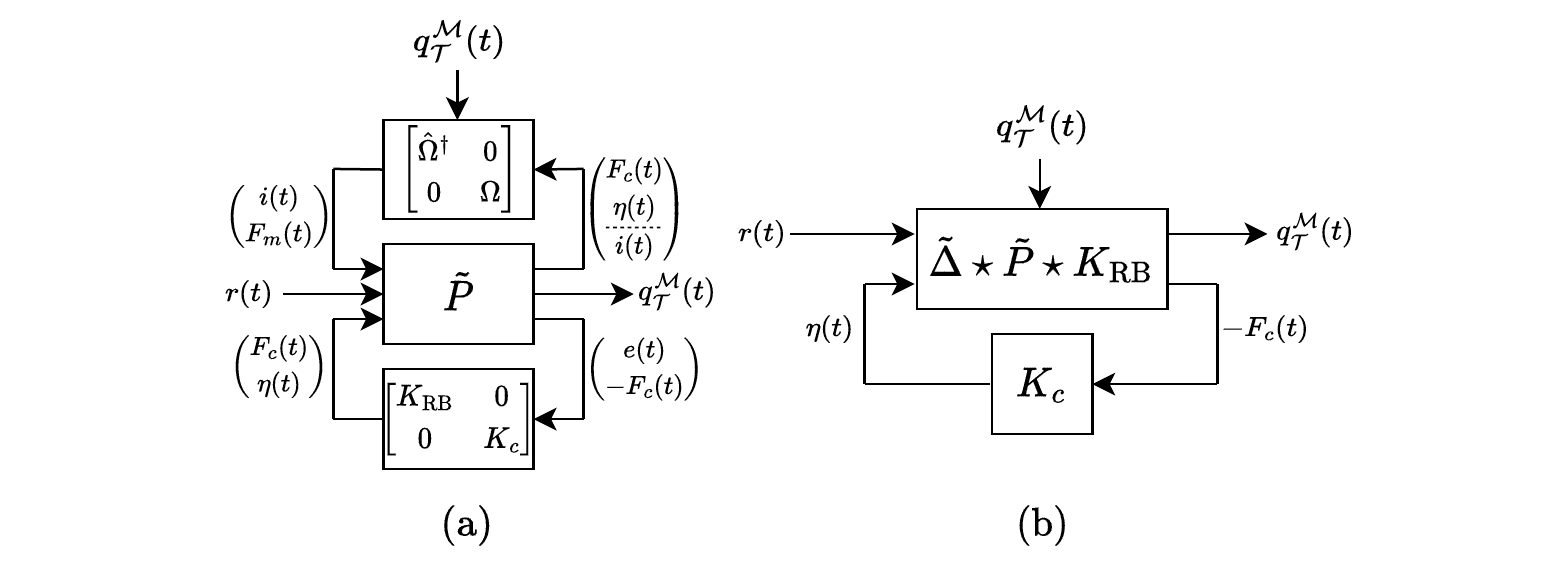}  \vspace*{-8mm}
\caption{(a): LFR representation of the control interconnection in Figure \ref{Fig:ContolInterconnection1}. (b): nonlinear plant as seen by the commutation controller $K_c$.}
\label{fig:LFR_Representation}
\end{figure}

In this Section, a novel and enhanced commutation approach is introduced that incorporates a dynamic regulation of the commutation frame. This regulation enables the adaptation of the commutation frame to account for local variations of the EM interaction\YB{, i.e. attenuation of the spatial behavior of $\Delta(q_{\mathcal{T}}^{\mathcal{M}})$}. Figure \ref{Fig:ContolInterconnection1} illustrates the proposed dynamic regulation, in which the commutation controller $K_c$ is designed to regulate the \YB{low-frequency} content of the rigid-body feedback control forces to zero, such that the objective function, expressed by (\ref{costfunctionpar2}), is satisfied \YB{for all $t \in [0\ \infty]$}. To allow for design of $K_c$, the control interconnection illustrated by Figure \ref{Fig:ContolInterconnection1} is reformulated as a \YB{\emph{linear fractional representation} (LFR)}, see Figure \ref{fig:LFR_Representation}(a), where the controllers $K_{\mathrm{RB}}$ and $K_c$ are extracted into a lower fractional connection while $\Omega$ and $\Omega^\dagger$ are lifted out as an upper connection, forming a diagonal block nonlinearity, which is denoted by $\tilde{\Delta}$. The resulting disconnected plant $\Tilde{P}$ is assumed to be a generalized plant\YB{, i.e. there exists at least one controller $\mathcal{K} =\text{diag}(K_{\mathrm{RB}},K_c)$ that stabilizes the open-loop interconnection $\tilde{P}$, see \cite{zhou1998essentials}.} Moreover, under the assumption that $K_{\mathrm{RB}}$ robustly internally stabilizes the plant in the conventional rigid-body control case, e.g., under $K_c = 0$, see Figure \ref{fig:LFR_Representation}(b), the resulting open loop transfer $r \rightarrow q_{\mathcal{T}}^\mathcal{M}$ is stable with respect to the block diagonal nonlinearity $\Tilde{\Delta}$ for the full operating envelope of the system. In this context, a set of local dynamics, denoted by $\lbrace  \tilde{\Delta} \star \tilde{P} \star K_{\mathrm{RB}} \rbrace_{i=1}^{\YB{N}}$, can be obtained by performing open-loop identification for various forced equilibria, i.e. \YB{around various operating points ${\tt q_i}$}, of the nonlinear system\YB{, where $\star$ is the star product.} \YB{By considering a dense sampled set of local dynamics, local stability around any forces equilibria ${\tt q_i}$ can be guaranteed by designing a LTI commutation controller $K_c$ using robust sequential loop closing strategies, see \cite{Oomen,skogestad2007multivariable}.} 

Analysis of experimental data collected from a cutting-edge MMPA prototype leads to several noteworthy observations. First, it is evident that, owing to the utilization of rigid body decoupling techniques in the position control loop, all local models are decoupled at low frequencies. Secondly, the diagonal elements of the local models demonstrate a 0 $\frac{\mathrm{dB}}{\mathrm{dec}}$ slope at low frequencies, which can be attributed to the static correlation between the compensatory control forces of the rigid body feedback controller and the relative position of the commutation frame. Therefore, the diagonal elements of the commutation controller are structured as PI-type of controllers, such that the steady-state error of the rigid body feedback control forces is actively regulated to zero by the integral action of the commutation controller.
Hence, the commutation feedback controller $K_c$ \YB{is} chosen \YB{as}:
 \vspace*{-1.5mm}
\begin{equation}
    K_c = \text{diag}\left(\frac{2\pi f_{bw}^i}{s} \right), \quad i \in [1 \ \ n_{{\eta}}],
    \label{PICOTROLLER}
\end{equation}

 \vspace*{-1.5mm}
\noindent where $s$ is the complex frequency and $f_{bw}^i$ denotes the intended commutation feedback control bandwidth, which is intentionally chosen to be 100 times smaller than the \emph{bandwidth of the position controller} to avoid excessive interaction between the two loops at high frequencies.

\YB{The suggested dynamic regulation approach meets three criteria. First, the design of the dynamic regulator $K_c$ utilizes FRF data, eliminating the need for extensive remodeling of EM relationships, thereby satisfying (R1). Secondly, the integral action of $K_c$ effectively mitigates any EM discrepancy, fulfilling (R2). Lastly, local stability is ensured through robust control design of $K_c$, stabilizing all local plants $\lbrace \tilde{\Delta} \star \tilde{P} \star K_{\mathrm{RB}} \rbrace_{i=1}^{N}$ and satisfying (R3). This dynamic compensation scheme contributes as C2 in the paper.}
 \vspace*{-1mm}
\section{Feedforward enhanced dynamic regulation}
\label{Section:CommutationFeedforward}

\begin{figure}[t]
\vspace{7pt}
    \centering
\includegraphics[trim={.5cm 0.5cm 2.2cm 0cm}
    ,width=\linewidth]{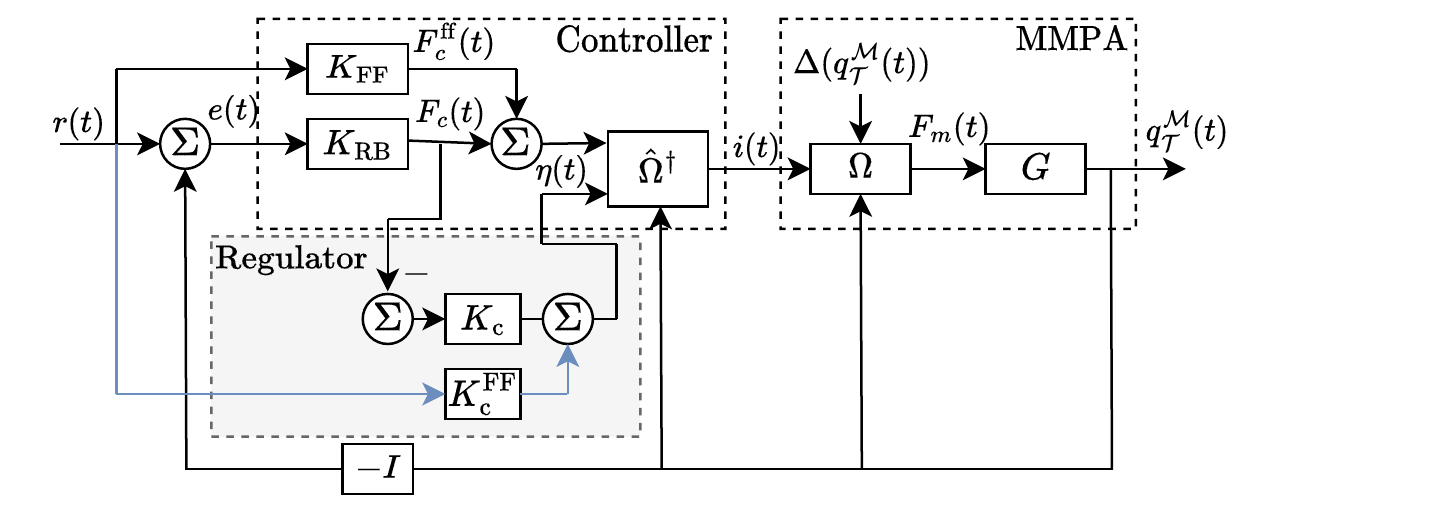}
    \caption{Proposed dynamic regulation of the commutation frame using commutation feedback controller $K_c$ and feedforward controller $K_c^{\mathrm{FF}}$.}
    \label{fig:fullcontrolllopp}
    \vspace*{-4mm}
    \vspace{-7pt}
\end{figure}

\begin{figure}[b]
\vspace*{-3mm}
    \centering  \includegraphics[trim={1.5cm 0.5cm 2.5cm 1.5cm}
    ,width=\linewidth]{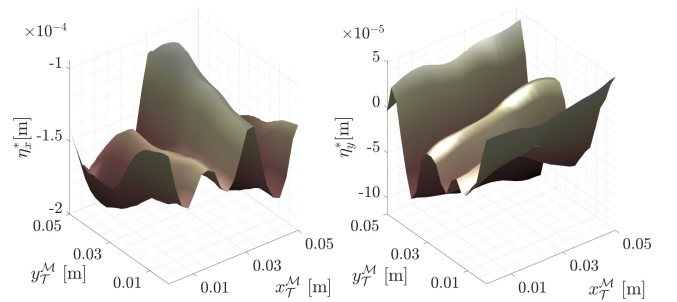}
    \caption{\YB{Optimized $\lbrace {\eta}_{j}^*(i) \rbrace_{i=1}^\mathrm{N}$ in terms of its $x$ and $y$ components, obtained from experiments using a state-of-the-art MMPA prototype, see \cite{proimadis2021active}.}}
    \label{fig:OptimalEta}
\end{figure}
In order to enhance the performance of the dynamic regulation of the commutation frame with a more aggressive controller, a commutation feedforward controller $K_c^{\mathrm{FF}}$ is introduced to the control interconnection, as depicted in Figure \ref{fig:fullcontrolllopp}. The design of the commutation feedforward can be approached in various ways. First, the solution of the gradient descent-based method, explained in section \ref{Section_GRADDESC}, can be employed, \YB{giving only a} static \YB{compensation} of the commutation frame. 
Alternatively, one can aim at constructing a position dependent feedforward policy that attempts to capture the spatial characteristics of  $\Delta(q_{\mathcal{T}}^{\mathcal{M}}(t))$ in terms of learning the mapping of $q_{\mathcal{T}}^{\mathcal{M}} \rightarrow \eta$. To construct such a feedforward,
$\YB{N}$ experiments are conducted with the aim of constructing a dataset $\mathcal{D}_N$, which consists of locally optimized parameter vectors as a function of the position, i.e.  $\mathcal{D}_N = \{ \eta^*(i),x_{\mathcal{T}}^{\mathcal{M}} (i),y_{\mathcal{T}}^{\mathcal{M}}(i)\}_{i=1}^N$. Moreover, this dataset is obtained by assessing the output of the dynamic regulator $K_c$ at $N$ local positions, \YB{e.g. using experiments with set-point control of a given $x_{\mathcal{T}}^{\mathcal{M}},y_{\mathcal{T}}^{\mathcal{M}}$ position}, which is facilitated by the integral action of the commutation controller. 
Figure \ref{fig:OptimalEta} illustrates the experimental findings in terms of its $\eta_x^*$ and $\eta_y^*$ components, revealing various noteworthy observations. Primarily, a static misalignment of the commutation frame is evident, as inferred by the non-zero mean of the \YB{obtained $\eta_x^\ast$, $\eta_y^\ast$ values}. Secondly, it is observed that the spatial behavior is dominated by nonlinear \YB{spatial} effects, which can be attributed to coil pitch, eddy currents, and other unmodelled EM relations.
To estimate the EM discrepancies $\Delta(q_{\mathcal{T}}^{\mathcal{M}})$ based on the outcomes of these local experiments, we adopt the \emph{Gaussian Process} (GP) framework, which offers a number of practical benefits compared to other techniques, \YB{such as} lookup tables and artificial neural networks. First, the GP framework provides confidence bounds on the posterior estimate. Additionally, the GP framework is able to \YB{accurately} model nonlinear mappings through appropriate kernel selection, which has been demonstrated to be effective in practical settings, see  \cite{proimadis2021learning, poot2022gaussian, VANMEER2022302}. Using the GP framework, the local variations of the EM relations are modelled as:
\vspace*{-1.5mm}
\begin{equation}
    \eta_\YB{j}^*(i) = \Delta_\YB{j}(w(i)) + \epsilon_\YB{j}(i), \quad \text{with} \ \epsilon_\YB{j} \sim \mathcal{N}(0,\sigma_{{\epsilon_{\YB{j}}}}^2),
\end{equation}

\vspace*{-1.5mm}
\noindent where $i$ corresponds to local position, $\eta_\YB{j}^*$ denotes the $\YB{j}$-th element of the $\eta^*$ vector and $\epsilon_\YB{j}$ is assumed to be independent and identically distributed (i.i.d.) white Gaussian noise with variance $\YB{\sigma_{{\epsilon_{\YB{j}}}}^2}$, see \cite{dekking2005modern}.  
The local commutation frame discrepancies $\Delta_\YB{j}$ are \YB{dependent on} an input vector $w(i)$, where $w(i)=[x_{\mathcal{T}}^{\mathcal{M}}(i) \ y_{\mathcal{T}}^{\mathcal{M}}(i)]^\top$, and are modelled as a GP following the approach presented in \cite{williams2006gaussian} as:
\vspace*{-1mm}
\begin{equation}
     \Delta_\YB{j}(w) \sim \mathcal{GP} (0,\kappa_\YB{j}(w,w'))
\end{equation}

\vspace*{-1mm}
\noindent where the covariance is fully characterized by the kernel functions $k_\YB{j}(w,w')$. Moreover, the joint Gaussian distribution of the observed target values $\{\eta_\YB{j}^* (i)\}_{i=1}^\YB{N}$ and the commutation frame discrepancies for a given test input $w^* \in \mathbb{R}^2$ is given by:
\vspace*{-1mm}
\begin{equation}
\resizebox{.89\hsize}{!}{$
    \begin{bmatrix}
        \lbrace{\eta}^*_\YB{j} (i)\rbrace_{i=1}^\YB{N} \\ \Delta_\YB{j}(w^*) 
    \end{bmatrix} \sim \mathcal{N}\left(0,\left [\begin{array}{cc}
        \kappa_\YB{j}(W,W) +\sigma_{{\epsilon_{\YB{j}}}}^2 I & \kappa_\YB{j}(W,w^*) \\
        \kappa_\YB{j}^\top(W,w^*) & \kappa_\YB{j}(w^*,w^*)
    \end{array}\right] \right),$}
    \label{jointdis}
\end{equation}

\vspace*{-1mm}
\noindent where $W \in {\mathbb{R}^{2n}}$ corresponds to \YB{the $x_{\mathcal{T}}^{\mathcal{M}}$, $y_{\mathcal{T}}^{\mathcal{M}}$ positions in }$\mathcal{D}_\YB{N}$ \YB{where measurements were made.} Based on $\mathcal{D}_\YB{N}$, the predictive a posteriori distribution computed from (\ref{jointdis}) is given by:
\vspace*{-2mm}
\begin{equation}
\resizebox{.85\hsize}{!}{$
    \begin{split}
    \hat{\Delta}_\YB{j}(w^*)&\triangleq  {{}\mathbb{E}}\{\Delta_\YB{j}(w^*) | \mathcal{D}_\YB{N},w^* \}=\Psi \YB{(w^\ast)} \\
       \mathrm{cov}\{ \Delta_\YB{j}(w^*) | \mathcal{D}_\YB{N},w^*\} &= \kappa_\YB{j}(w^*,w^*)-\Psi \YB{w^\ast} \kappa_\YB{j}(W,w^*)
    \end{split},
    \label{GPposterior}$}
\end{equation}

\vspace*{-1mm}
\noindent \YB{where $\Psi = \kappa_\YB{j}^\top(W,w^*)(\kappa_\YB{j}(W,W)+\sigma_{{\epsilon_{\YB{j}}}}^2 I)^{-1}$. }
Moreover, the construction of the feedforward controller $K_{c}^{\mathrm{FF}}$ involves using the mean of the \emph{posterior \YB{GP} distribution}\YB{, see (\ref{GPposterior}),} leading to a diagonal \YB{GP}-based feedforward control strategy:
\vspace*{-2mm}
\begin{equation}
    K_c^{\mathrm{FF}} = \mathrm{diag}(\{ \hat{\Delta}_\YB{j} (w^*) \}_{\YB{j}=1}^{n_\eta})
\end{equation}

\vspace*{-1mm}

\noindent where it is reasonable to assume that the test input is $w^* \approx [r_x \ r_y]^\top$. Based on this assumption, the \YB{GP}-based estimator can be employed as a feedforward controller, as depicted in Figure \ref{fig:fullcontrolllopp}. Achieving a precise and reliable estimator $\hat{\Delta}_j$ hinges on the careful selection of the kernels $k_j(w,w')$, which determine the \emph{Hilbert space} within which the mean of (\ref{GPposterior}) is searched for as a true estimate of $\Delta(q_{\mathcal{T}}^{\mathcal{M}})$. In light of the experimental findings presented in Figure \ref{fig:OptimalEta}, we propose the adoption of the following kernel selection:
\vspace*{-1mm}
\begin{equation}
\resizebox{.88\hsize}{!}{$
    k_\YB{j}(w,w') = \sigma_{\YB{j}_1}^2\text{exp} \Bigg(-\sum_{v=1}^2 \frac{||w_v-w_v'||_2^2}{\sigma^2_{\YB{j}_{v+1}}}- \sum_{v=1}^2 
    \frac{2\text{sin}^2\left(\frac{w_v-w_v'}{2} \right)}{\sigma^2_{\YB{j}_{v+3}}     } \Bigg)
  $},
    \label{kernel}
\end{equation}

\vspace*{-1mm}
\noindent 
where $w_v$ corresponds to the $v_{\mathrm{th}}$ element of the input vector, i.e. $[x_{\mathcal{T}}^\mathcal{M}$ \ $y_{\mathcal{T}}^\mathcal{M}]^\top$. The proposed kernel is composed of a combination of a periodic kernel \cite{mackay1998introduction} and a radial basis function kernel \cite{neal2012bayesian}. \YB{The periodic kernel is included to account for the sinusoidal effects in Figure \ref{fig:OptimalEta}, while the radial basis function kernel captures the non-linear residual EM effects.}
The proposed kernel functions, given by (\ref{kernel}), are characterized by 6 hyper-parameters, namely $[\sigma_{\YB{j}_1}^2 \ \hdots \ \sigma_{\YB{j}_5}^2 \ \sigma_{\epsilon_{{\YB{j}}}}^2]^\top$\YB{, which are tuned by the marginalized likelihood approach detailed in }\cite{rasmussen2006gaussian}. For validation of the \YB{obtained} GP \YB{estimates}, the {Best Fit Ratio} (BFR) is used, see \cite{simpkins2012system}, which is given by:
\vspace*{-1mm}
\begin{equation}
\resizebox{.85\hsize}{!}{$
    \mathrm{BFR} = 100 \% \cdot \mathrm{max}\left(1 - \frac{\left | \left | \lbrace  {{\eta}_{\YB{j}}^*(i)} \rbrace_{i=1}^{\YB{N}} - \lbrace \hat{\Delta}_{\YB{j}} (w(i)) \rbrace_{i=1}^{\YB{N}} \right | \right |_2}{\left |\left |\lbrace {{\eta}_{\YB{j}}^*(i)}  \rbrace_{i=1}^{\YB{N}} - { \Bar{{\eta}}_{\YB{j}}^*} \right |\right |_2} ,0\right)$},
\end{equation}

\vspace*{-1mm}
\noindent where \begin{small}${\Bar{{\eta}}_{\YB{j}}^*}$\end{small} corresponds to the sample mean of \begin{small}$\lbrace {{{\eta}}_{\YB{j}}^*(i)} \rbrace_{i=1}^{{\YB{N}}}$\end{small}. 
Table \ref{table:GPResults} presents the BFR results of the GP-based feedforward in terms of its $x$ and $y$ components, relative to both its training dataset $\mathcal{D}_{\YB{N}}$ and a novel validation dataset. The findings indicate that the proposed combination of a periodic kernel and a radial basis function kernel, enable the GP-based estimator to accurately capture the local discrepancies of the EM relationships, thus rendering it a reliable commutation feedforward controller. \YB{The GP-based enhanced dynamic regulation is Contribution C3 in the paper.}


\begin{table}[t]
\vspace{7pt}
\begin{center}
\caption{\YB{Validation of  the estimated GPs}}
\label{table:GPResults}
\scalebox{.9}{
\begin{tabular}{|c|c|c|}
\hline
             & \textbf{Training data set} & \textbf{Validation data set} \\ \hline
\textbf{BFR} $\hat{\Delta}_x (w^*)$ [\%] & 89.80                 & 85.18                   \\ \hline
\textbf{BFR} $\hat{\Delta}_y(w^*)$[\%]  & 84.84                & 83.34                   \\ \hline
\end{tabular}}
\end{center}
\vspace*{-6mm}
\vspace{-7pt}
\end{table}

\begin{figure}[b]
\vspace*{-3mm}
    \centering
\includegraphics[height=4cm,width=\linewidth]{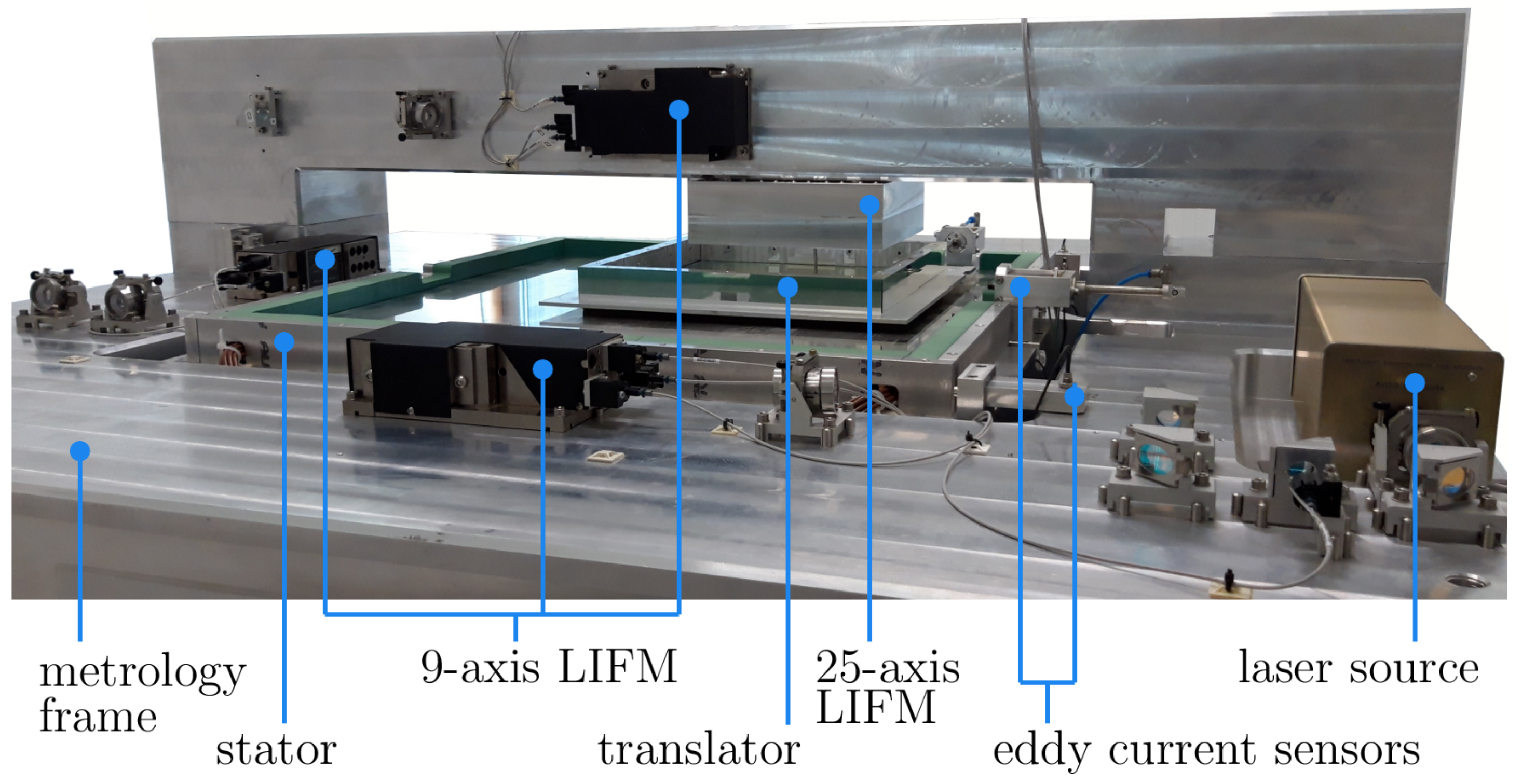}
    \vspace{-5mm}
    \caption{Photograph of a moving-magnet planar actuator system prototype.}
    \label{fig:MMPA}
\end{figure}

\vspace*{-.2mm}
\section{Experimental validation}
\label{Section_Experimental_Validation}
\vspace*{-1mm}
\subsection{MMPA prototype system}

The MMPA system, depicted in Figure \ref{fig:MMPA}, consists of three main components: a stator base on which a double layer coil array is mounted, a lightweight translator on which a Hallbach array, comprised of 281 permanent magnets, is mounted, and a metrology frame on which 9 laser interferometers are mounted to measure the relative displacement of the translator. The double layer coil array of the stator base consists of 160 coils of which 40 coils are simultaneously activated at every time instant using 40 power amplifiers, allowing for levitation and propulsion of the magnet plate in 6 DoF. 
For a more comprehensive description of the MMPA prototype, readers can refer to \cite{proimadis2021active}.


\vspace*{-.5mm}
\subsection{Experimental results}

To illustrate the functionality of the proposed approaches, i.e. static calibration of the commutation frame (Section \ref{Section_GRADDESC}) and GP-feedforward enhanced dynamic regulation of the commutation frame (Section \ref{Section_Dynamic_Regulation}-\ref{Section:CommutationFeedforward}), time-domain experiments have been performed on the MMPA prototype \YB{to compare the proposed approaches with the current state-of-the-art control configuration}. For implementation of the proposed approaches on the experimental prototype, the static EM calibration approach, presented in Section \ref{Section_GRADDESC}, was employed based on $\YB{N=}$ 36 local positions, leading to an optimal static commutation frame alignment parameter. For implementation of the GP-feedforward enhanced dynamic regulation of the commutation frame, the GP-based feedforward, discussed in Section \ref{Section:CommutationFeedforward}, is implemented in conjunction with the dynamic regulation proposed in Section \ref{Section_Dynamic_Regulation}, \YB{where the GP is trained using 576 datapoints.} The commutation feedback controllers are designed based on (\ref{PICOTROLLER}), \YB{with a }bandwidth 100 times smaller than the bandwidth of the corresponding controllers of the position loop\YB{, i.e. $f_{\mathrm{bw}}^i \approx 1$-$2 Hz$.} \YB{Note that the linear controller $K_c$ has been robustly designed to stabilize all local plants $\lbrace  \tilde{\Delta} \star \tilde{P} \star K_{\mathrm{RB}} \rbrace_{i=1}^{\YB{N}}$.}

\begin{figure}[t]
\vspace*{7pt}
    \centering
    \includegraphics[trim={1.5cm 0cm 2.5cm 0cm}
    ,width=.95\linewidth,height=4cm]{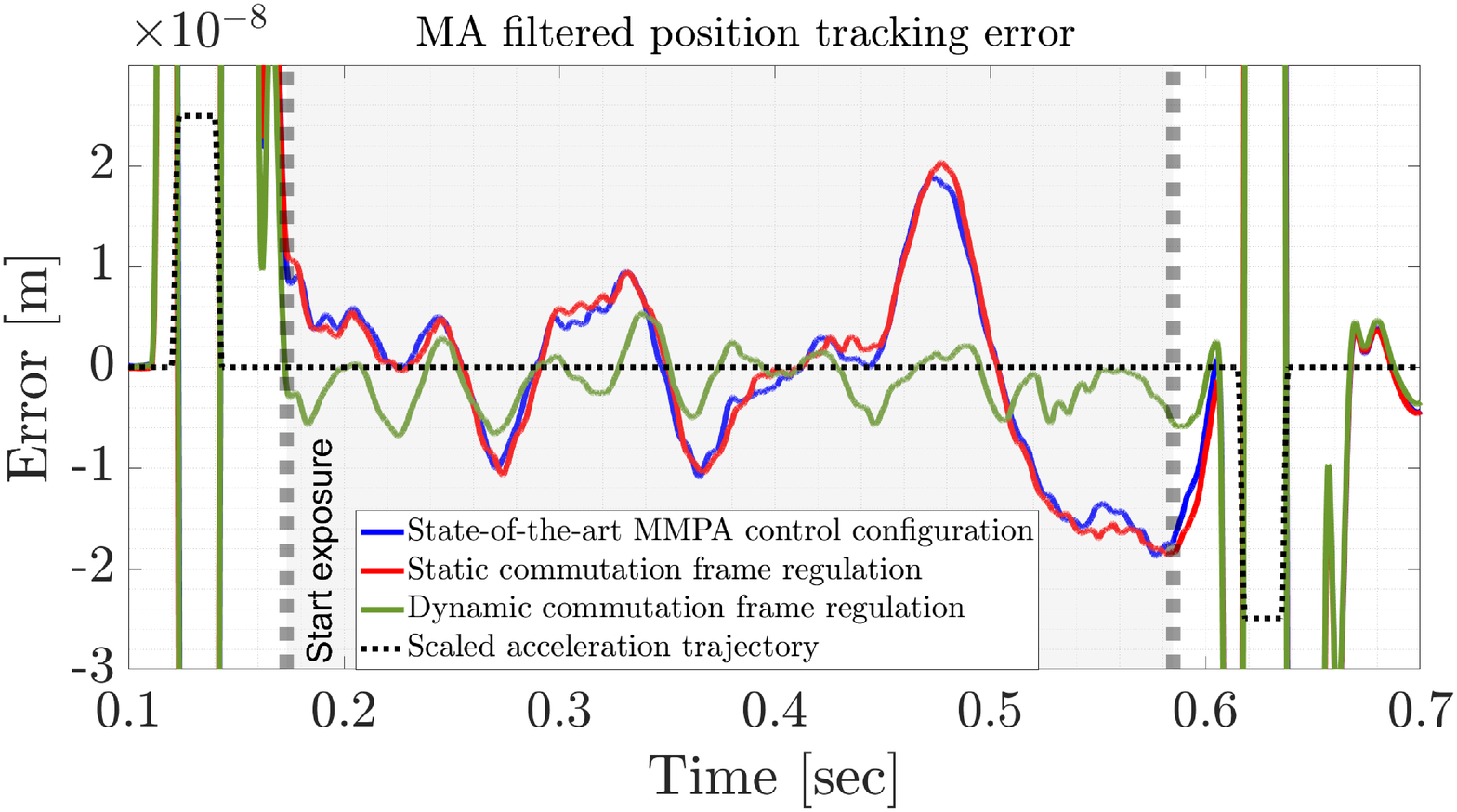}
    \vspace{-3mm}
    \caption{\YB{MA filtered position tracking error in $y$-direction during constant velocity interval of the motion profile (grey region) with: (\Large\textcolor{blue}{-}\footnotesize) Standard commutation, (\Large\textcolor{red}{-}\footnotesize) Static alignment of the commutation frame (Section \ref{Section_GRADDESC}) (\Large\textcolor{green}{-}\footnotesize)  Dynamic regulation of the commutation frame (Section \ref{Section_Dynamic_Regulation}) combined with a learning-based commutation feedforward (Section \ref{Section:CommutationFeedforward}).}}
    \label{fig:EXPResults}
    \vspace*{-4mm}
\end{figure}

To assess performance in lithographic applications, the Moving-Average (MA) performance metric is introduced, see \cite{Butler}, which allows to analyse the \YB{low-frequency} spectral content of the position tracking error during the lithographic exposure process, which takes place in the constant velocity interval of the motion profile, \YB{which will be indicated with the color grey}. The MA performance metric is given by:
\vspace*{-1mm}
\begin{equation}\begin{small}
        \mathrm{MA}(t) = \frac{1}{T}\int_{t-\frac{T}{2}}^{t+\frac{T}{2}} e(\tau)d\tau,
\end{small}
        \label{MAFILTER}
\end{equation}

\vspace*{-1mm}
\noindent where $e$ corresponds to the raw position tracking error and $T=0.0144s$ is the exposure time of a single point on the wafer. For experimental validation of the proposed approaches, a lithographic scanning motion is performed in the $y_{\mathcal{T}}^{\mathcal{M}}$ direction using a fourth order motion profile, where $a_{\mathrm{max}}$ = $5 \frac{m}{s^2}$, $v_{\mathrm{max}}$ = $0.1\frac{m}{s}$, and $x_{\mathrm{max}}$ = $0.05{m}$. The experimental results, e.g. \YB{{the MA filtered position tracking errors in $y_{\mathcal{T}}^{\mathcal{M}}$-direction during the constant velocity interval of the motion profile}}, are illustrated in Figure \ref{fig:EXPResults}. Specifically, the blue graph depicts the position tracking error in  of the \YB{current state-of-the-art} control structure for MMPAs, the red graph \YB{illustrates} the position tracking error using the proposed static EM alignment approach \YB{on top of the current control structure} and the green graph corresponds to the position tracking error of the proposed GP-feedforward enhanced dynamic regulation of the commutation frame \YB{on top of the current control structure}.
\YB{Based on Figure \ref{fig:EXPResults}, the static calibration approach (Section \ref{Section_GRADDESC}) is insufficient for improving position tracking due to the nonlinear spatial behavior of $\Delta(q_{\mathcal{T}}^{\mathcal{M}})$ caused by coil pitch, eddy currents and manufacturing imperfections, see Figure \ref{fig:OptimalEta}. However, implementation of the GP-feedforward enhanced dynamic regulation of the commutation frame reduces the peak MA error by 64.71\%, from approximately 18.7 nm to 6.6 nm, demonstrating the potential of this approach for enhancing the EM relationships in the model-based commutation.}

\vspace*{-1mm}
\section{Conclusions}
\label{Section_Conclusions}

\YB{This paper introduces three methods to enhance the model-based commutation in planar motors by regulating the commutation frame statically and dynamically to address EM discrepancies from misalignment, eddy currents, coil pitch, and manufacturing imperfections. Experimental results on a MMPA prototype demonstrate that static alignment alone does not sufficiently improve position tracking performance due to spatial EM behavior. However, the proposed GP-based feedforward enhanced dynamic regulation reduces peak MA error by 64.71\% compared to conventional commutation.}

\addtolength{\textheight}{-12cm}   



\vspace*{-1mm}
\bibliographystyle{ieeetr}       
\bibliography{MyBib}

\end{document}